\documentclass[seceq]{ptptex}

\usepackage{graphicx}

\title{
Precise asymptotics for a variable-range hopping model%
}

\author{
Bernhard  \textsc{Mehlig}$^{1)}$ and Michael \textsc{Wilkinson}$^{2)}$%
}

\inst{
 $^{1}$Department of Physics, G\"oteborg University,
41296 Gothenburg, Sweden\\$^{2}$Faculty of Mathematics and
Computing, The Open University, Walton Hall, Milton Keynes, MK7
6AA, England 
}

\abst{
For a system of localised electron states the DC conductivity
vanishes  at zero temperature, but localised electrons can conduct
at finite temperature. Mott gave a theory for the low-temperature
conductivity in terms of a variable-range hopping model, which is
hard to analyse. Here we give precise asymptotic results for a
modified variable-range hopping model, proposed by S.
Alexander [{\it Phys. Rev. B}, {\bf 26}, 2956 (1982)].
}

\begin{document}

\maketitle

\section{Introduction}
If a system has localised electron states
the DC conductivity must be zero at zero temperature, but
localised electrons can conduct at finite temperatures. Mott
\cite{Mot79} proposed that the low-temperature dependence of the
DC conductivity is
\begin{equation}
\label{eq: 1} \sigma(T) \sim
\exp\Big[-\Big(\frac{T_0}{T}\Big)^\frac{1}{d+1}\Big]
\end{equation}
in $d$ spatial dimensions  (variable-range hopping law).

\begin{figure}[t]
\includegraphics[width=7cm,clip]{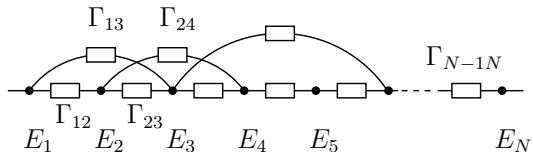}
\caption{\label{fig:0} One-dimensional resistor-network model for
a one-dimensional variable-range hopping problem. The conductances
$\Gamma_{nm}$ are given by (\ref{eq: 2}), $E_n$ are random on-site
energies.}
\end{figure}

The variable-range hopping process is described by specifying
transition rates for an electron to hop from one localised site to
another. The standard model is illustrated in one spatial
dimension in figure \ref{fig:0}. It is of the form of a resistor
network, with nodes labelled by $n=1,\ldots,N$. The conductances
$\Gamma_{nm}$ are given by the transition rates
\begin{equation}
\label{eq: 2} \Gamma_{nm}=\Gamma_0\,
\exp\left(-\frac{\displaystyle |E_n| + |E_m| + |E_n-E_m|}
{\displaystyle 2 k_{\rm B} T}\right)\, \epsilon^{\displaystyle
\vert n-m\vert}
\end{equation}
where $E_m$ are random on-site energies (measured relative to the
Fermi energy). Here the exponential factor accounts for the
temperature dependence of the rate constants and the factor
$\epsilon^{\vert n-m\vert}$ models the decrease of the matrix
elements as a function of distance between the sites. The matrix
elements are assumed to decrease exponentially as a function of
hopping distance, with parameter $\epsilon \equiv \exp(-\alpha)$
(where $\alpha$ is the inverse localisation length). In the
low-temperature limit the rate constants differ by orders of
magnitude, so that the conduction of the chain is determined by a
single path, and the conductivity is the harmonic mean of the
dominant transition rates. Transitions have a higher rate if they
require a small increase in the energy of the electron, which
favours finding a conduction path involving low-energy states.
There is thus a competition between hopping to a nearby site,
which is favoured by large transition matrix elements, and finding
an energetically favoured site, which may be some distance away.

Despite the appealing simplicity of this physical picture, there
are no precise asymptotic expressions and few significant results
on variable-range hopping. Ambegaokar {\em et al.} \cite{Amb71}
have employed the fact that the transition rates between localised
electron states in disordered solids fluctuate wildly to map the
variable-range hopping problem onto a percolation problem. This
explains the exponent in (\ref{eq: 1}) for $d > 1$ and provides an
estimate of $T_0$. Kurkij\" arvi \cite{Kur73} showed that the DC
conductivity in infinitely long disordered chains is simple
activated conduction, following the Arrhenius law: $\sigma(T) \sim
\exp(-T_0/T)$. This result was later refined by Raikh and Ruzin
\cite{Rai89}. In one dimension, Mott's law (\ref{eq: 1}) is
expected to be valid for sufficiently short chains only.

Alexander \cite{Ale82} introduced a simplified model for variable-range
hopping. In this paper we discuss the physical significance
of his model and analyse it using an alternative method. We obtain
a precise asymptotic expression for the conductivity according to
this model.

\section{The Alexander model}
 Finding the true conduction path is
a difficult problem. Here, following Ref. [5], we adopt a
simpler approach. We construct a conduction path, visiting a
sequence of sites by the following scheme. Given the $i$th site in
the sequence, at position $n$ in the lattice, we take the site
$i+1$ to be the site with position $m$ which has the largest
transition rate $\Gamma_{nm}$ [given by equation (\ref{eq: 2})],
considering only sites lying to the right. (That is, we maximise
$\Gamma_{nm}$ with respect to $m$, subject to $m>n$). This
approach will not usually produce the optimal conduction path, for
which a non-optimal transition at one step may be more than
compensated by a lower overall resistance. However, the model has
the appealing feature that its asymptotics can be determined
precisely, as we now show.

We can simplify the model further by assuming that the rate is
$\Gamma_{nm}=\epsilon^{\vert n-m\vert-\omega_n/t}$, where
$\omega_n$ is a random number and $t$ a small parameter.
In the following we assume that the $\omega_n$ are independently,
identically uniformly distributed in $[0,1]$.

\section{Method of solution}
Because the values of $\Gamma_{nm}$ differ by orders of magnitude
we can approximate the largest rate $\Gamma$ from a given site to
some other site by the sum  over all rates
\begin{equation}
\label{eq: 3} \Gamma=\sum_{n=0}^\infty \epsilon^n x_n
\end{equation}
where $x_n={\rm e}^{-\omega_n/t}$.
The probability density for the random variables $x_n$ is
\begin{equation}
\rho(x)=\left\{
\begin{array}{ll}
t/x &  \mbox{for}\,\quad \exp(-t^{-1})\leq x \leq 1\\
0 & \mbox{otherwise.}
\end{array}\right .
\end{equation}
The conductivity of the system is obtained from the diffusion
constant $D$, which is given by $D=-J\Delta N/\Delta p$, where
$\Delta p$ is the change in occupation probability over $\Delta N$
sites. We have $\Delta p=J\sum_i \Gamma_i^{-1}=J\langle
\Gamma^{-1}\rangle \Delta N_{\rm step}$, where $\Delta N_{\rm
step}$ is the number of maximum-rate hops in the process described
above. If the maximum-rate hops have average length $\langle
\Delta n\rangle$, then $\Delta N_{\rm step}=\Delta N/\langle
\Delta n\rangle$, and the diffusion constant is
\begin{equation}
\label{eq:D} D=\langle \Gamma^{-1}\rangle^{-1}\,      \langle
\Delta n\rangle \ .
\end{equation}
The quantity $\langle\Delta n\rangle$ may be determined by
considering
\begin{equation}
\epsilon {\partial \log \Gamma\over{\partial \epsilon}}={\sum_n
n\epsilon^n x_n\over{\sum_n \epsilon^n x_n}}\,.
\end{equation}
This will almost always be approximately equal to the value of $n$
for which the rate $\epsilon^n x_n$ is maximal. Thus we may write
\begin{equation}
\label{eq:4} \langle \Delta n \rangle=\epsilon
{\partial\over{\partial \epsilon}} \langle \log \Gamma \rangle
\end{equation}
and we have the ingredients necessary to determine $D$ from the
probability density $\rho_\Gamma$ of $\Gamma$.

\begin{figure}[t]
\includegraphics[width=10cm,clip]{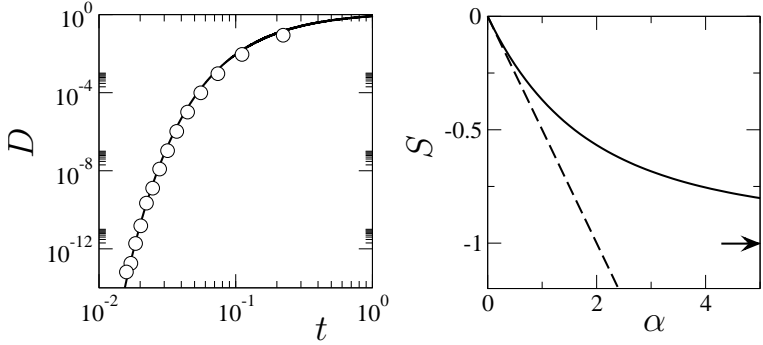}
\caption{\label{fig:1}
  (Left) shows results of
  numerical simulations for diffusion constant $D$
  as a function of $t$ ($\circ$) for $\epsilon = 0.2$, compared
  to the exact asymptotic theory, equation (\ref{eq:result}).
  (Right) shows $S(\alpha)$, equation (\ref{eq:S}), as a function of $\alpha$
  (solid line). Also shown is the approximate result $S(\alpha) \sim
  -\alpha/2$ from Ref. [5] (dashed line). The arrow indicates the
  nearest-neighbour limit $S = -1$.}
\end{figure}

\section{Distribution of $\Gamma$}
The distribution of $\Gamma$
may be obtained from $\rho(x)$ [the distribution of the variables
$x_n$ in (\ref{eq: 3})], by application of the convolution
theorem. Because $\Gamma$ is a sum of independent random
variables, its probability density is the convolution of the
probability density of each component, namely
$\rho(x/\epsilon^n)$. Correspondingly, the generalised Fourier
transform $\tilde \rho_\Gamma (k)$ of the required probability
density is a product:
\begin{equation}
\label{eq:14}
\tilde \rho_\Gamma (k)=\prod_{n=0}^\infty \tilde \rho(k
\epsilon^n)
\end{equation}
where the generalised Fourier transform is defined by
\begin{equation}
\tilde \rho(k)=\int_{-\infty}^\infty\!\! {\rm d}x\ \exp({\rm i}kx)\rho
(x)\,.
\end{equation}
Setting $u=kx$  we obtain
\begin{eqnarray}
\tilde \rho (k)&=&t
\int_{\hspace*{-8mm}\raisebox{-3mm}{$\scriptstyle k\exp(-1/t)$}}^k
\hspace*{-4mm}
{\rm d}u\ u^{-1}\exp({\rm i}u)\\[0.2cm]
\nonumber &=&t\bigl[E_1(-{\rm i}k\exp(-1/t)-E_1(-{\rm i}k)\bigr]
\end{eqnarray}
where $E_1(v)$ is the exponential integral
\begin{equation}
E_1(v)=\int_x^\infty {\rm d}z\ {\exp(-z)\over z}\,.
\end{equation}
From the asymptotic form for $E_1(v)$ we find $\tilde \rho(0)=1$
(which is required by normalisation).
Further we find $\vert \tilde \rho(k) \vert
< t$ for $k>\exp(1/t)$, and that for $1 \ll k \ll \exp(1/t)$ we can
apply the following approximation:
\begin{equation}
\tilde \rho(k)\sim 1-t\log k -\gamma t+\tfrac{1}{2}{\rm i}\pi t\,.
\end{equation}
We now determine $\tilde \rho_\Gamma
(k)$. In the case where $\vert k\vert\ll 1$ all of the factors
are close to unity such that $\tilde \rho_\Gamma (k)\sim
1$ in that region. In the case where $\vert k\vert \gg \exp(1/t)$,
the product is negligible, and we approximate $\tilde \rho_\Gamma
(k)\sim 0$ in that region. Finally, when $1 \ll k \ll \exp(1/t)$,
most of the factors in the infinite product are very close to
unity, and we can replace the infinite sum by a finite number $N^\ast$ of
terms. This number is determined by the value $N^\ast$
of the index such that the argument of the corresponding term,
$\epsilon^{N^\ast}k$ is equal to unity. Thus $N^\ast$ is the integer part
${\log k}/{\log(1/\epsilon)}={\log k}/{\alpha}$
and we have
\begin{equation}
\tilde \rho_\Gamma (k)\sim\prod_{n=0}^{{\rm Int}[(\log k)/\alpha]}
\big(1-t\log \epsilon^n k -\gamma t +\tfrac{1}{2}\pi {\rm
i}t\big)\,.
\end{equation}
Evaluating the product  we obtain an approximate expression
for $\tilde \rho_\Gamma (k)$, valid for $t\ll 1$ and $1\ll k\ll \exp(1/t)$:
\begin{eqnarray}
\label{eq:17}
\tilde \rho_\Gamma (k)&\sim & (\alpha t)^{1+\alpha^{-1}\log k}\,\,
\frac{\Gamma\displaystyle\Big(\frac{1+\alpha t-\gamma t+ {\rm i}\pi t/2}{\alpha t}\Big)}
     {\Gamma\displaystyle\Big(\frac{1-t\log k- \gamma t+ {\rm i}\pi t/2}{\alpha t}\Big)}\,.
\end{eqnarray}
We need to determine the asymptotic form for small values of $t$.
Note that $x = \log k$  varies between $0$ and $t^{-1}$ so
that $xt$ is of order unity. Using
\begin{equation}
\label{eq:18}
\Gamma(z+a) \sim \sqrt{2\pi}\, {\rm e}^{-z} z^{-\frac{1}{2}+z+a}
\end{equation}
for large $z$ and taking $z = (1-xt)/(\alpha t)$,
we obtain
\begin{eqnarray}
\nonumber
\tilde \rho_\Gamma(k) &=& {\rm e}^{\displaystyle -\frac{x}{\alpha}
+\big(\frac{1}{2}-\frac{1-xt}{\alpha t}+\frac{\gamma}{\alpha}
-\frac{{\rm i}\pi}{2\alpha}\big) \log(1-xt)}\,.\\
\end{eqnarray}
The calculation of $D$ requires two averages:   $\langle
\Gamma^{-1}\rangle$ and $\langle \log \Gamma \rangle$. Averages
can be calculated in terms of the generalised Fourier transform of the
probability density:
\begin{equation}
\langle f(\Gamma)\rangle = \int_{-\infty}^\infty {\rm d}\Gamma\
\rho(\Gamma) f(\Gamma)=\frac{1}{2\pi}\int_{-\infty}^\infty {\rm
d}k\ \tilde \rho_\Gamma (k)\tilde f(k)
\end{equation}
where $\tilde f(k)$ is the generalised Fourier transform of the
weight function $f(\Gamma)$.

\section{Harmonic average of $\Gamma$}
We have
\begin{equation}
\langle \Gamma^{-1}\rangle
=\int_0^\infty {\rm d}k\  \mbox{Im}[\tilde\rho_\Gamma (k)]
\end{equation}
since the generalised Fourier transform of $1/\Gamma$ is
$\pi {\rm i} \,\mbox{sgn}(k)$.
We evaluate the integral in the saddle-point approximation.
To this end let $y=xt = t\log k$. Note that  the contribution to
integrals with respect to $k$ from the interval $[0,1]$ is
negligible and is not considered.
We thus have
\begin{equation}
\langle \Gamma^{-1}\rangle \sim \frac{1}{t} \int_0^1 \!{\rm d}y\,
{\rm e}^{t^{-1}S(y)} {\rm e}^{(\frac{1}{2}
+\frac{\gamma}{\alpha})\log(1-y)} \sin\phi(y)
\end{equation}
with
\begin{eqnarray}
\label{eq: 21}
S(y) &=& y-\frac{y}{\alpha} -\frac{1-y}{\alpha }\log(1-y)\,,\\
\phi(y) &=& \frac{\pi}{2\alpha}\log(1-y)\,.
\nonumber
\end{eqnarray}
The saddle-point equations are:
\begin{eqnarray}
y^\ast &=& 1-{\rm e}^{-\alpha}\,,\\
S(y^\ast) &=& \frac{-1+\alpha+{\rm e}^{-\alpha}}{\alpha}\,,\nonumber\\
\phi(y^\ast) &=& -\frac{\pi}{2}\,,\nonumber\\
\frac{\partial^2S^\ast}{\partial y^2} &=&-\frac{{\rm
e}^\alpha}{\alpha}\,.
\nonumber
\end{eqnarray}
It follows that
\begin{equation}
\label{eq:hm} \langle \Gamma^{-1}\rangle  \sim
\sqrt{\frac{2\pi\alpha}{t}}\, {\rm e}^{\displaystyle
-\alpha-\gamma } \,{\rm e}^{\displaystyle \frac{-1+\alpha+ {\rm
e}^{-\alpha}}{\alpha t}}\,.
\end{equation}

\section{Average of $\log\Gamma$}
 To  calculate $\langle \log
\Gamma\rangle$, we make use of the relation
\begin{equation}
\label{eq:32}
\langle \log \Gamma \rangle=\frac{2}{\pi}
\int_0^\infty {\rm d}k\ {\log
\vert k
\vert +\gamma \over k} \mbox{Im}[\tilde \rho_\Gamma (k)]\,.
\end{equation}
From equation (\ref{eq:32}) we find
\begin{eqnarray}
\langle \log\Gamma\rangle &\sim& \frac{1}{t^2} \frac{2}{\pi}
\int_0^1\!\!\!{\rm d}y\, (y+\gamma t)\,{\rm e}^{(\alpha t)^{-1} R(y)} 
 {\rm e}^{(\frac{1}{2}
 +\frac{\gamma}{\alpha})\log(1-y)}\sin\phi(y)\,.
\end{eqnarray}
with
\begin{eqnarray}
R(y) &=& -y-(1-y)\log(1-y)
\end{eqnarray}
and $\phi$ as defined in (\ref{eq: 21}).
The integral is approximately
\begin{equation}
\langle \log\Gamma\rangle \sim -\frac{1}{t^2} \frac{2}{\pi}
\frac{\pi}{2\alpha} \int_0^1\!\!\!{\rm d}y\, y(y+\gamma t)\,{\rm
e}^{(\alpha t)^{-1} R(y)}\,.
\end{equation}
Approximating $R(y) = -y^2/2$, we obtain
\begin{equation}
\langle \log \Gamma \rangle \sim -(\pi/2)^{1/2}(\alpha/t)^{1/2}\,.
\end{equation}

\section{Results}
Using (\ref{eq:4}) it follows that
\begin{equation}
\label{eq:42} \langle \Delta n \rangle \sim \frac{1}{4}
\sqrt{\frac{2\pi}{\alpha t}}\,.
\end{equation}
Taking equations (\ref{eq:D}), (\ref{eq:hm}), and (\ref{eq:42})
together we finally obtain:
\begin{equation}
\label{eq:result} D \sim \frac{1}{4\alpha} {\rm e}^{\alpha+\gamma}
{\rm e}^{[1-\alpha-\exp(-\alpha)]/(\alpha t)}\,.
\end{equation}
This is the main result of this paper. It describes the exact
asymptotics of the diffusion constant in Alexander's model,
exhibiting Arrhenius behaviour, with action (excitation energy)
given by
\begin{equation}
\label{eq:S}
S(\alpha) = \frac{1-\alpha-\exp(-\alpha)}{\alpha}
\end{equation}
which is plotted in Fig. \ref{fig:1}.
The limit  $S(0) =  -\alpha/2$ (dashed line in Fig. \ref{fig:1})
was previously obtained in Ref. [5].
In the opposite limit ($\alpha\rightarrow \infty$)
the action converges to $-1$, describing activated transport
in a conventional nearest-neighbour model.

In addition to the precise form (\ref{eq:S}) of the action, our
result (\ref{eq:result}) also gives the pre-exponential prefactor
exactly. We know of no variable-range hopping model for which the
pre-exponential factor has been obtained precisely. The precise
asymptotic expression for $D$ allows for comparison with computer
simulations. The simulations were performed as follows: we sampled
the rate $\Gamma$ for the optimal hop from a given point to the
right, as well as the corresponding length $\Delta n$. The
diffusion constant was then calculated from (\ref{eq:D}). In
figure \ref{fig:1}, equation (\ref{eq:result}) is shown together
with results of computer simulations for $\epsilon = 0.2$
(corresponding to $\alpha \approx  1.62$). We observe excellent
agreement between the simulations and the precise asymptotic
theory.

{\em Acknowledgements}. BM acknowledges financial support
from Vetenskapsr\aa{}det.

\end{document}